\documentclass[10pt,conference]{IEEEtran}
\IEEEoverridecommandlockouts
\usepackage{cite}
\usepackage{amsmath,amssymb,amsfonts}
\usepackage[T1]{fontenc}
\usepackage{graphicx}
\usepackage{textcomp}
\def\BibTeX{{\rm B\kern-.05em{\sc i\kern-.025em b}\kern-.08em
    T\kern-.1667em\lower.7ex\hbox{E}\kern-.125emX}}

\usepackage{ulem}
\usepackage[dvipsnames]{xcolor}

\usepackage[switch]{lineno}
\usepackage[colorlinks= false, urlcolor=blue]{hyperref}
\hypersetup{
    colorlinks=true,            
    citecolor=purple,
    linkcolor=Green,
    urlcolor=Blue            
}
\usepackage{booktabs}

\usepackage{listings, xcolor}
\definecolor{verylightgray}{rgb}{.97,.97,.97}
\lstdefinelanguage{Solidity}{
  keywords=[1]{anonymous, assembly, assert, balance, break, call, callcode, case, catch, class, constant, continue, constructor, contract, debugger, default, delegatecall, delete, do, else, emit, event, experimental, export, external, false, finally, for, function, gas, if, implements, import, in, indexed, instanceof, interface, internal, is, length, library, log0, log1, log2, log3, log4, memory, modifier, new, payable, pragma, private, protected, public, pure, push, require, return, returns, revert, selfdestruct, send, solidity, storage, struct, suicide, super, switch, then, this, throw, transfer, true, try, typeof, using, value, view, while, with, addmod, ecrecover, keccak256, mulmod, ripemd160, sha256, sha3}, 
  keywordstyle=[1]\color{blue}\bfseries,
  keywords=[2]{address, bool, byte, bytes, bytes1, bytes2, bytes3, bytes4, bytes5, bytes6, bytes7, bytes8, bytes9, bytes10, bytes11, bytes12, bytes13, bytes14, bytes15, bytes16, bytes17, bytes18, bytes19, bytes20, bytes21, bytes22, bytes23, bytes24, bytes25, bytes26, bytes27, bytes28, bytes29, bytes30, bytes31, bytes32, enum, int, int8, int16, int24, int32, int40, int48, int56, int64, int72, int80, int88, int96, int104, int112, int120, int128, int136, int144, int152, int160, int168, int176, int184, int192, int200, int208, int216, int224, int232, int240, int248, int256, mapping, string, uint, uint8, uint16, uint24, uint32, uint40, uint48, uint56, uint64, uint72, uint80, uint88, uint96, uint104, uint112, uint120, uint128, uint136, uint144, uint152, uint160, uint168, uint176, uint184, uint192, uint200, uint208, uint216, uint224, uint232, uint240, uint248, uint256, var, void, ether, finney, szabo, wei, days, hours, minutes, seconds, weeks, years},  
  keywordstyle=[2]\color{teal}\bfseries,
  keywords=[3]{block, blockhash, coinbase, difficulty, gaslimit, number, timestamp, msg, data, gas, sender, sig, value, now, tx, gasprice, origin},  
  keywordstyle=[3]\color{violet}\bfseries,
  identifierstyle=\color{black},
  sensitive=false,
  comment=[l]{//},
  morecomment=[s]{/*}{*/},
  commentstyle=\color{gray}\ttfamily,
  stringstyle=\color{red}\ttfamily,
  morestring=[b]',
  morestring=[b]"
}
\usepackage{pifont}
\usepackage{diagbox}
\usepackage{subfigure}

\usepackage{color}
\usepackage{microtype}

\usepackage{oplotsymbl}
\usepackage{subfigure}
\usepackage{siunitx}
\usepackage{array,framed}
 \usepackage{
   color,
   float,
   epsfig,
   wrapfig,
   graphics,
   graphicx,
 }
 \usepackage{xcolor} 
\usepackage{setspace}
\usepackage{amsfonts}
\usepackage{latexsym,fancyhdr,url}
\usepackage{enumerate}
\usepackage{algorithm2e}
\usepackage{algpseudocode}
\usepackage{graphics}
\usepackage{xparse} 
\usepackage{xspace}
\usepackage{multirow}
\usepackage{microtype}
\usepackage[most]{tcolorbox}

\usepackage{soul}
\soulregister\cite7 
\soulregister\ref7 

\usepackage{color}

\usepackage{fancyhdr}

\usepackage{subfigure}
\newcommand{\para}[1]{\vspace{2pt}\noindent\textbf{#1.~}}

\newcommand{\CollectedDApp}{576}
\newcommand{\DAppGT}{54}

\newcommand{\WildDApp}{835}
\newcommand{\WildHasOne}{459}
\newcommand{\OverallRecall}{84.06\%}
\newcommand{\OverallPrecision}{92.10\%}

\begin{document}

\title{Hyperion: Unveiling DApp Inconsistencies using LLM and Dataflow-Guided Symbolic Execution}

\author{
    \IEEEauthorblockN{Shuo Yang$^{\dag}$, Xingwei Lin$^{\ddag}$, Jiachi Chen\thanks{* Jiachi Chen is the corresponding author.}$^{\dag}$\IEEEauthorrefmark{1}, Qingyuan Zhong$^{\dag}$, Lei Xiao$^{\dag}$, Renke Huang$^{\dag}$, \\Yanlin Wang$^{\dag}$, Zibin Zheng$^{\dag}$}
    \IEEEauthorblockA{$^\dag$ Sun Yat-sen University, \{yangsh233, zhongqy, xiaolei6, huangrk9\}@mail2.sysu.edu.cn}
     \IEEEauthorblockA{\{chenjch86, wangylin36, zhzibin\}@mail.sysu.edu.cn}
    \IEEEauthorblockA{$^\ddag$ Zhejiang University, xwlin.roy@zju.edu.cn}
}

\maketitle

\begin{abstract}
The rapid advancement of blockchain platforms has significantly accelerated the growth of decentralized applications (DApps).
Similar to traditional applications, DApps integrate front-end descriptions that showcase their features to attract users, and back-end smart contracts for executing their business logic.
However, inconsistencies between the features promoted in front-end descriptions and those actually implemented in the contract can confuse users and undermine DApps's trustworthiness.

In this paper, we first conducted an empirical study to identify seven types of inconsistencies, each exemplified by a real-world DApp.
Furthermore, we introduce \textsc{Hyperion}, an approach designed to automatically identify inconsistencies between front-end descriptions and back-end code implementation in DApps.
This method leverages a fine-tuned large language model LLaMA2 to analyze DApp descriptions and employs dataflow-guided symbolic execution for contract bytecode analysis. Finally, \textsc{Hyperion} reports the inconsistency based on predefined detection patterns. The experiment on our ground truth dataset consisting of 54 DApps shows that \textsc{Hyperion} reaches 84.06\% overall recall and 92.06\% overall precision in reporting DApp inconsistencies. We also implement \textsc{Hyperion} to analyze 835 real-world DApps. The experimental results show that \textsc{Hyperion} discovers 459 real-world DApps containing at least one inconsistency.
\end{abstract}


\section{Introduction}
In recent years, the blockchain industry has witnessed a remarkable proliferation of various decentralized applications (DApps), which have gained substantial popularity and market capitalization~\cite{wu2019empirical, werner2022sok, nftdefects}. 
Similar to traditional applications like web or Android apps, a DApp utilizes a user interface (UI) as its front-end to present feature descriptions and employs smart contracts for back-end logic execution.

However, inconsistency between the front-end descriptions and the actual implementation of their back-end contract code of DApps may have detrimental consequences~\cite{duan2022towards}.
For example, a DApp might advertise a 3\% investment return in its description, but the actual return could be changed (lower than 3\%) after investment.
Similarly, in the case of some NFT DApps, the front-end description may claim that NFTs can ``live forever'', while in reality, the metadata of these NFTs is stored on centralized servers, which are susceptible to shutdown. These inconsistencies can pose threats to users' interests and undermine the trustworthiness of DApps.  

Many research efforts have been made to detect vulnerabilities in smart contracts to ensure their safety~\cite{luu2016making,nftdefects,tsankov2018securify,choi2021smartian,jiang2018contractfuzzer}.
However, they usually overlook the importance of inconsistencies between the front-end descriptions and back-end implementations in DApps. For instance, consider a scenario where a DApp offers users a 2\% interest rate and does not mention fees; traditional vulnerability detection tools may not flag this as an issue. However, when the promised rates in the DApp's description do not align with this or fees are charged secretly, it can indicate dishonest behavior of the DApps' owners and pose potential risks to DApp users.

The detection of DApp inconsistencies presents unique challenges.
First, there is no systematic work aimed at uncovering the various types of inconsistencies within the DApp ecosystem. This absence makes it not easy to design rules to detect inconsistencies. Second, the diverse and intricate structures of DApp front-end descriptions present considerable difficulties when attempting to extract information related to user assets.
Third, it is challenging to recover the semantics from the contract code effectively. Many malicious DApp developers may not disclose their source code on Ethereum~\cite{liao2022smartdagger}. Consequently, detecting inconsistencies from the bytecode level is crucial, but also increases the difficulty.

To address these challenges, we first conducted an empirical study employing an open card sorting approach~\cite{spencer2009card} to find DApp inconsistencies. This involved analysis of both the front-end descriptions and the smart contracts of a dataset comprising 321 DApps (see Section~\ref{sec:define}). Based on this approach, we define seven types of DApp inconsistencies that encompass both DeFi and NFT DApps, namely \textit{Unguaranteed Reward}, \textit{Hidden Fee}, \textit{Adjustable Liquidity}, \textit{Unconstrained Token Supply}, \textit{Unclaimed Fund Flow}, \textit{Changeable DApp Status}, and \textit{Volatile NFT Accessibility}.

To automatically detect these inconsistencies, we propose a tool named \textsc{Hyperion}, which contains a description analyzer \textsc{HyperText} and a contract analyzer \textsc{HyperCode}.
\textsc{HyperText} leverages LLaMA2~\cite{touvron2023llama}, a widely-recognized, public-available large language model~\cite{chen2023chatgpt} (LLM), to scrutinize the diverse natural language descriptions found on DApp websites. 
To optimize LLaMA2 for the specific domain task, i.e., extract critical attributes from the DApp front-end descriptions, we first design a prompt template based on the chain of thought~\cite{wei2023chainofthought} (CoT) prompting technique, tailored to our identified inconsistency types. Then, we manually label a dataset containing 63 DApp inconsistencies to fine-tune LLaMA2 (see Section~\ref{sec:fine-tune}), thereby enhancing its performance for our downstream task and obtaining our model \textsc{HyperText}. Finally, we employ the Natural Language Tool Kit~\cite{nltk} (NLTK) to extract inconsistency-related attributes from \textsc{HyperText}'s output (see Section~\ref{sec:nltk}).




For the analysis of DApp contract bytecode, \textsc{HyperCode} employs an approach that utilizes dataflow analysis to guide symbolic execution on the contract intermediate representation (IR), so as to obtain relevant program states and recover contract semantics (see Section~\ref{sec:sca}).
Specifically, \textsc{HyperCode} first decompiles the contract bytecode and performs dataflow analysis to recover low-level semantics such as contract call and state variable storage access operations.
Then, it performs a graph analysis based on the recovered semantics to obtain the contract's fund transfer and state variable dependency relationship.
To obtain contract semantics, \textsc{HyperCode} proposes a symbolic execution framework based on contract IR. This framework utilizes dataflow and graph analysis to optimize search paths and direct symbolic execution to check relevant program states, merging the strengths of both methods.
Furthermore, we propose algorithms that compare the information extracted from DApp descriptions and contract semantics based on this framework to detect the seven defined inconsistencies, unveiling whether one DApp's \textit{saying} is consistent with its \textit{doing}.

To evaluate the performance of \textsc{Hyperion}, we first collect two datasets; one serves as the ground truth dataset, containing  \DAppGT~labeled DApps used to define DApp inconsistencies and evaluate the effectiveness of the \textsc{Hyperion}. The other dataset is used to evaluate the performance of \textsc{Hyperion} on analyzing real-world wild DApps, which contains \WildDApp~real-world DApps collected from DappBay~\cite{DappBay} and DappRadar~\cite{DappRadar}.
\textsc{Hyperion} reaches an overall recall of \OverallRecall~and an overall precision of 92.06\% in the first ground truth dataset. Furthermore, \textsc{Hyperion} identifies \WildHasOne~out of \WildDApp~DApps with at least one inconsistency that we define in the second large-scale dataset, with an overall precision of \OverallPrecision~in our sampled dataset.
Among the true positives (268 DApps) in the sampled dataset, we find that 67 (25\%) DApps' websites become inaccessible within only 3 months.


In summary, the contributions of our paper are as follows:
\begin{itemize}
\item We define seven common inconsistency types between DApp front-end description and back-end smart contract implementation. To enhance comprehension, we illustrate each inconsistency with a real-world DApp example and its consequences.
\item We design and develop \textsc{Hyperion}, a tool that leverages LLM and dataflow-guided symbolic execution to automatically detect defined inconsistencies. 
\item We find \WildHasOne~out of \WildDApp~DApps in our dataset contain at least one inconsistency, which demonstrates the prevalence of the defined inconsistencies in the DApp ecosystem.

\item To promote further research and transparency, we have released the source code of \textsc{Hyperion} (both fine-tuned \textsc{HyperText} and \textsc{HyperCode}) as well as the DApp dataset resources and experimental results in the repository {\href{https://github.com/shuo-young/Hyperion}{https://github.com/shuo-young/Hyperion}}.

\end{itemize}



\vspace{-0.2cm}
\section{Background}
\subsection{DApp and Smart Contract}
Decentralized applications (DApps) typically leverage smart contract technology to achieve autonomous and transparent operations~\cite{zhou2023dapphunter,zheng2020overview}.
A DApp usually uses a user interface (UI) that showcases its functionalities to attract users. Some financial-related features, such as return on investment (ROI), token supply, and liquidity, are central to user interest and are implemented within the smart contract. Smart contracts are commonly written in high-level programming languages, e.g., Solidity~\cite{solidity}. Ethereum Virtual Machine (EVM) is a stack-based virtual machine that executes the contract EVM bytecode.
The EVM executes transactions by splitting contract EVM bytecode into operation codes (opcodes), each with specific execution instructions.

\vspace{-0.2cm}
\subsection{Large Language Model}\label{sec:llmintro}
Recently, Large language models (LLMs) have exhibited remarkable proficiency in natural language understanding~\cite{icse24-llm}. 
Notably, ChatGPT~\cite{chen2023chatgpt} is distinguished by its advanced performance. However, it faces limitations such as limited availability in some regions and high API usage fees. 
In contrast, LLaMA~\cite{touvron2023llama} offers advantages in terms of transparency, adaptability, cost-free access, and robust performance in natural language tasks. These features make LLaMA an ideal model for complicated or unstructured natural language understanding.
Additionally, the ability of LLaMA in downstream tasks can be further enhanced by its fine-tuning capabilities, a process of training the model on specific task-related labeled data in a supervised manner~\cite{devlin2018bert,liu2016recurrent}. This adaptability is further amplified in the context of instruction-tuning~\cite{zhang2023instruction,ouyang2022training,li2023tuna}, which trains models to interpret and act on explicit instructions in diverse tasks.
Numerous models have been developed based on LLaMA, including Code LLaMA~\cite{roziere2023code}, optimized for understanding and generating programming code, and LLaMA2~\cite{llama2}, which offers enhanced capabilities for natural language processing. Given our specific objective of comprehending intricate natural language data in DApp descriptions, we select LLaMA2 as our base model.

\vspace{-0.2cm}
\section{Inconsistencies Definition}\label{sec:define}
\begin{figure}[htbp!]
\setlength{\abovecaptionskip}{-0.1cm}
\setlength{\belowcaptionskip}{-0.5cm}
    \centering
    \includegraphics[width=3.3in]{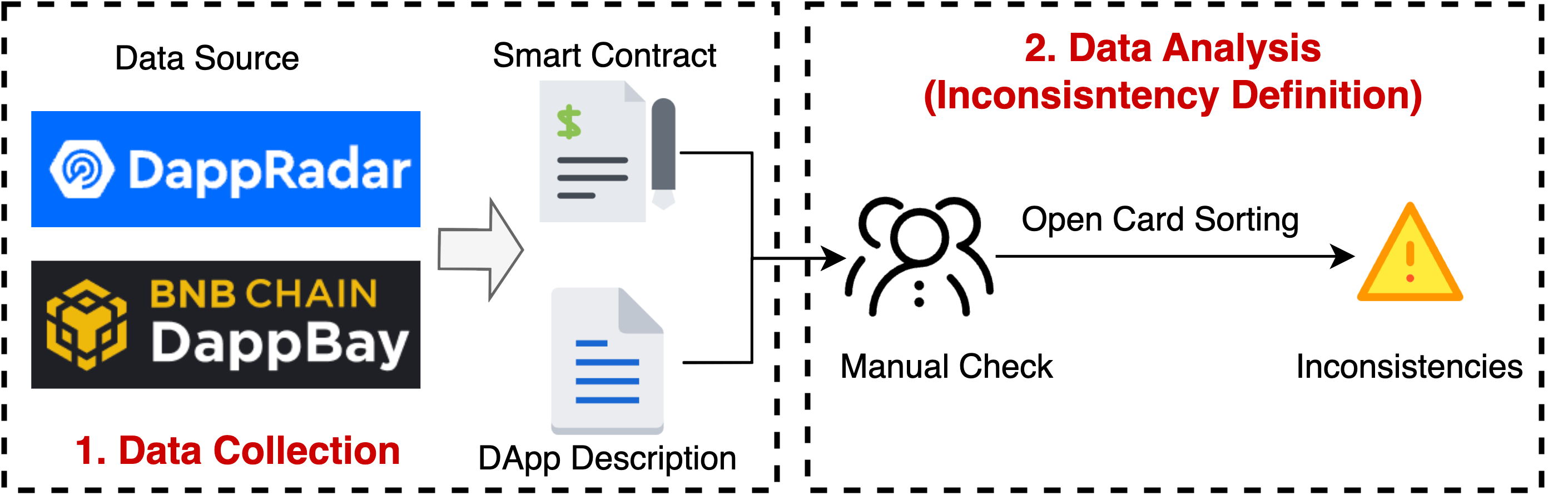}
    \caption{Workflow of defining DApp inconsistencies.}
    \label{fig:data}
\end{figure}

Figure~\ref{fig:data} illustrates how we define the seven types of DApp inconsistencies. First, we collected DApp descriptions and corresponding smart contracts from two platforms. Then, we employed the open card sorting approach~\cite{spencer2009card} to analyze the collected data manually. Finally, we defined seven DApp inconsistencies, each illustrated with a real-world DApp example.

\subsection{Data Collection}

We first crawled all \CollectedDApp~DApps labeled as \textit{high-risk} or \textit{red-alarm} by DappRadar~\cite{DappRadar} and DappBay~\cite{DappBay}, two main platforms that offer detailed information on DApps.
We then removed 255 DApps whose websites are inaccessible for further analysis and manually collected the HTML of each DApp by visiting their official websites. Next, we extracted related smart contract addresses from these websites and collected their source codes from Etherscan.
We finally obtained descriptions and smart contracts of 321 DApps. 

\vspace{-0.3cm}

\vspace{-0.3cm}
\subsection{Data Analysis}\label{sec:opencardsoring}

To find DApp inconsistencies, three authors use the open card sorting approach~\cite{spencer2009card}, which is widely adopted in problem-finding and definition in software engineering~\cite{nftdefects,chen2020defining}.
In this process, we first created a card that contains the front-end description and the contract code for each DApp. We considered two aspects to ensure the significance of the inconsistency issues, namely user financial-related information in DApp descriptions (e.g., promised return on investment, claimed fees charged, and the stability of the token economy) and corresponding financial-related code logic in smart contract implementations. These aspects are highly related to users' profits or assets and are usually the major concern of users. Then, three authors worked together to determine the labels for each card. They followed the detailed steps illustrated in the previous problem definition works~\cite{nftdefects,chen2020defining}.

There are three steps in the open card sorting process. In the first step, they randomly chose 40\% of the cards and read the DApp descriptions. They extracted the claims that are related to users' funds or assets, e.g., \textit{our DApp pays 8\% daily for users}, or \textit{we charge another 4\% fee for marketing}, etc.
Similarly, they read the contract code and focused on the implementations related to the fund transfer or tokens. They checked the transfer target, the transfer amount, and other attributes related to the assets. 
For those contracts without source code, they used an online decompiler~\cite{Dedaub} and analyzed the contract IR.
Based on the two aspects of information extracted from the DApp description and the smart contract, they compared them and understood the inconsistencies.

In the second step of card sorting, the same three authors analyzed and categorized the remaining 60\% cards independently, following the same steps as in the first round. 
Meanwhile, if they encountered new descriptions or contract implementation related to users' funds or assets other than what they found in the first round (e.g., daily reward), they kept the cards to the final discussion.

In the third step, they compared the results, discussed the differences, and finally identified seven types of inconsistencies across \DAppGT~DApps from all the collected \CollectedDApp~DApps. The remaining 267 DApps found no further inconsistencies or had no financial-related codes or descriptions during the open card sorting process.
We established the mapping relationship between the filtered DApp and the defined inconsistencies, which is publicly available in our repository.

\subsection{Definition of DApp Inconsistencies}
Table~\ref{tab:inconsistency} shows the definition of seven inconsistency types between the DApp front-end descriptions and smart contracts. The third column highlights the inconsistency types, and to clearly show each inconsistency, we list both the illustration of front-end descriptions and the implementation of smart contracts (columns fourth and fifth). Our defined inconsistencies cover NFT and DeFi DApps.

The following paragraphs provide the corresponding detailed definition and example for each inconsistency.

\begin{table*}[htbp]
\setlength\tabcolsep{3pt}
    \setlength{\abovecaptionskip}{-0.5cm}
    \setlength{\belowcaptionskip}{-0.7cm}
    \begin{center}
        \caption{Definitions of the 7 DApp Inconsistencies.}
        \label{tab:inconsistency}
\resizebox{\textwidth}{!}{%
\begin{tabular}{l|l|l|l|l}
\hline
\textbf{General Type}                  & \textbf{DApp Type}                 & \textbf{Inconsistency Type}  & \textbf{Front-end Description}                                               & \textbf{Smart Contract Implementation}                           \\ \hline
\multirow{2}{*}{Mathematics}  & \multirow{3}{*}{DeFi}     & \textit{Unguaranteed Reward (UR)}              & Claim a reward of x\%                                               & Reward rate is not a guaranteed constant value                     \\ \cline{3-5} 
                              &                           & \textit{Hidden Fee (HF)}                 & Not claim fee or claim fee as x\%                                & Fees are charged with a rate of y\% (y!=x)                    \\ \cline{1-1} \cline{3-5} 
\multirow{5}{*}{Transparency}  &                           & \textit{Adjustable Liquidity (AL)}        & Claim x years of liquidity lock time                                           & The lock time are adjustable to be less than x                      \\ \cline{2-5} 
                              & \multirow{3}{*}{DeFi/NFT} & \textit{Unconstrained Token Supply (UTS)} & Claim a token supply of x                                &    The supply can be bigger than x                        \\\cline{3-5}  &                           & \textit{Unclaimed Fund Flow (UFF)}           & Not claim the possible fund flow to the project owner            & The contract fund can be withdrawn by specific users \\ \cline{3-5} 
                              &                           & \textit{Changeable DApp Status (CDS)}         & Not claim the possible pause of the DApp                         & The DApp can be paused to impede trading                \\ \cline{2-5} 
                              & NFT                       & \textit{Volatile NFT Accessibility (VNA)}   & Not claim the accessibility of NFTs or claim they are accessible & The NFTs can be inaccessible                            \\ \hline

\end{tabular}%
}
\end{center}
\vspace{-0.4cm}
\end{table*}

\textbf{(1) Unguaranteed Reward (UR):}
Rewards are pivotal in incentivizing user investment in DApps.
Some DApps advertise high rewards to attract investments, but inconsistencies between these advertised rewards and the contract's actual implementation can deceive users. Such inconsistencies might lead to users not receiving the guaranteed rewards or even resulting in financial loss.

\vspace{-0.3cm}
\begin{figure}[htbp]
\setlength{\abovecaptionskip}{-0.1cm}
\setlength{\belowcaptionskip}{-0.3cm}
    \centering
    \includegraphics[width=3.3in]{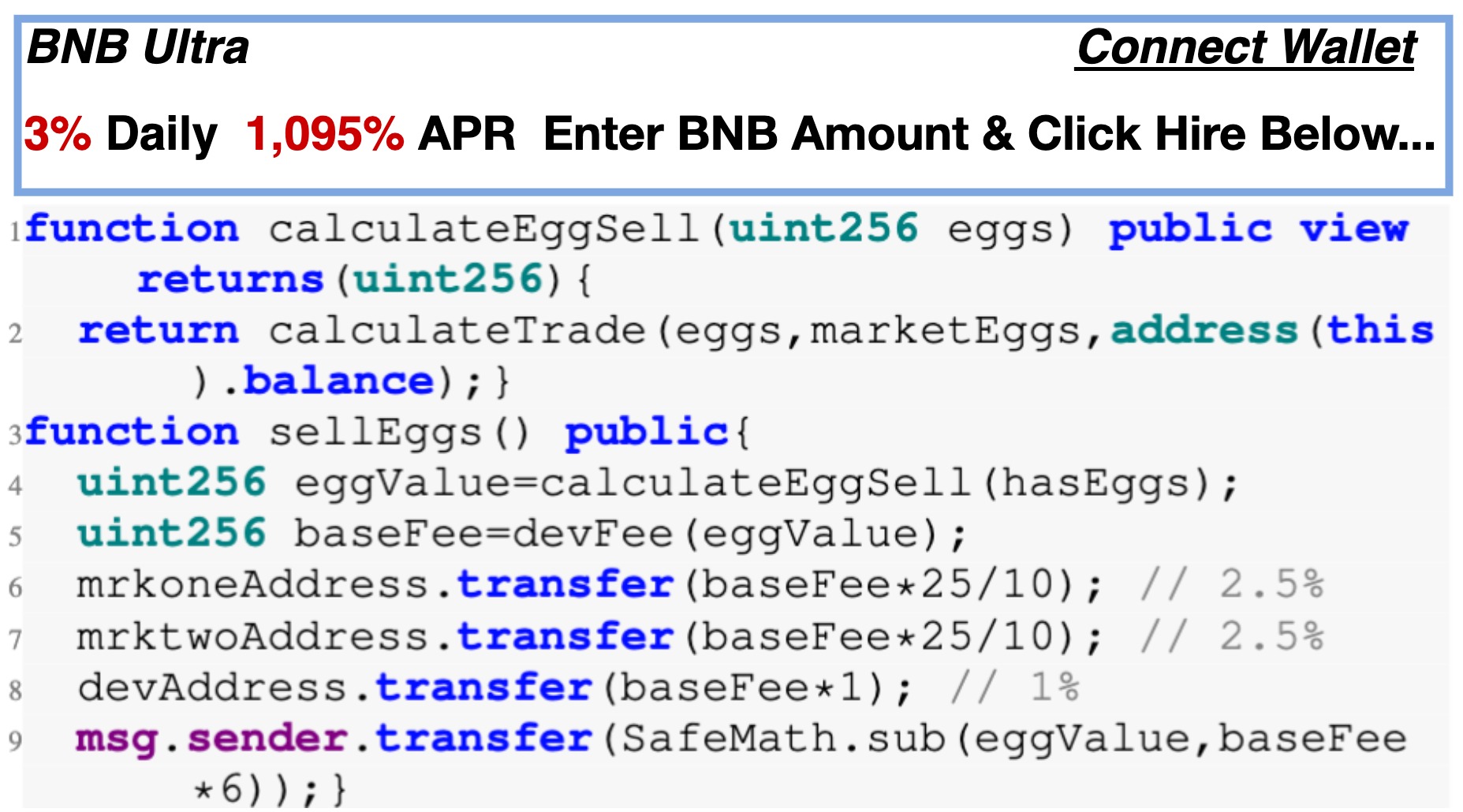}
    \caption{DApp descriptions (top) and contract snippet (bottom) of \textit{BNB Ultra}.}
    \label{fig:reward}
\end{figure}
\vspace{-0.1cm}

\textbf{Example:} Descriptions on \textit{BNB Ultra}~\cite{bnbultra} (Figure~\ref{fig:reward}) might mislead users that they can obtain a daily profit of 3\%. 
An examination of the contract~\cite{bnbultracontract} reveals that the reward is not guaranteed. 
Specifically, line 9 indicates that the user reward comprises two components, i.e., the value of \textit{eggs} and \textit{baseFee}. The value of \textit{eggs} is determined by the function \textit{calculateTrade()} (line 2), depending on variable \verb|marketEggs| and contract balance. This implementation means that the promised 3\% reward rate is subject to undetermined variable factors, contradicting the front-end claim.

\vspace{-0.3cm}
\begin{figure}[!htbp]
\setlength{\abovecaptionskip}{-0.1cm}
\setlength{\belowcaptionskip}{-0.5cm}
    \centering
    \includegraphics[width=3.3in]{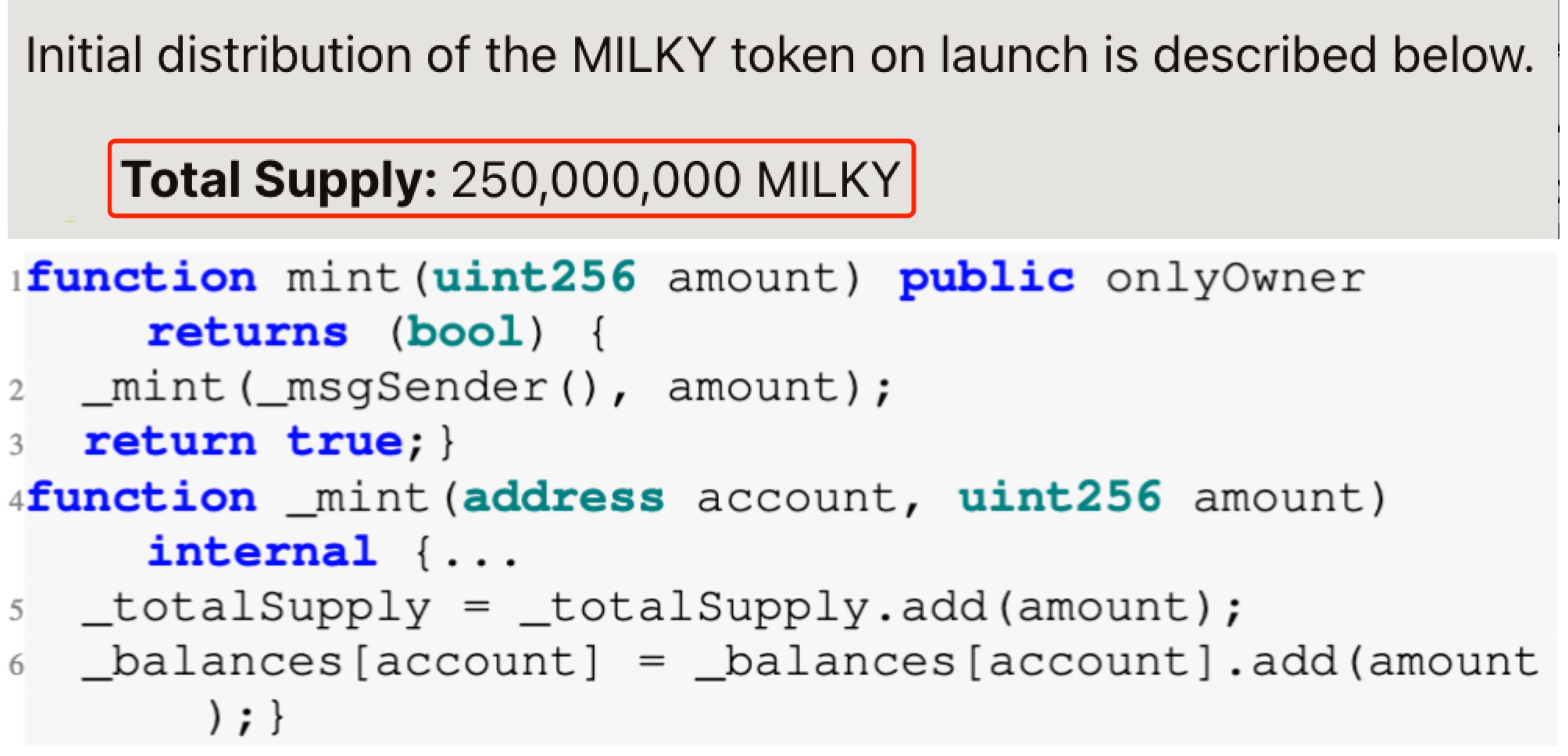}
    \caption{DApp description (top) and contract snippet (bottom) of \textit{MILKY} token.}
    \label{fig:supply}
\end{figure}
\vspace{-0.3cm}

\textbf{(2) Hidden Fee (HF):}
DApps often charge fees to maintain their operations, compensating for resources such as management efforts. However, issues arise when these fees are either undisclosed on the DApp's front-end description or differ from the actual contract implementation, which can unexpectedly impact user assets and compromise the DApp's trustworthiness.


\textbf{Example:} 
Returning to \textit{BNB Ultra}~\cite{bnbultra}, the final reward for users is reduced by fees directed to predetermined addresses (Figure~\ref{fig:reward}, line 9). Specifically, lines 6-8 show fees being allocated to three different addresses, with the total fee amount determined by the \textit{devFee()} function (line 5). However, this fee structure~\cite{bnbultracontract} is not disclosed in the DApp’s front-end description (Figure~\ref{fig:reward}), resulting in a \textit{Hidden Fee} inconsistency. 

\textbf{(3) Unconstrained Token Supply (UTS):}
Many DApps have their native tokens, e.g., Aave~\cite{aave} and BAYC~\cite{bayc}. Setting the maximum supply of tokens can uphold tokens scarcity~\cite{nftdefects}, stabilizing the market.
Inconsistencies between the stated token supply and actual contract code can detrimentally affect token holders by affecting the token's value and scarcity~\cite{brekke_digital_2021}.


\textbf{Example:}
Descriptions about the \textit{MILKY} token~\cite{bpmliky} claims a total supply of 250M (Figure~\ref{fig:supply}). However, the contract~\cite{mlikycontract} (line 2) allows one to mint tokens without a supply cap. 


\textbf{(4) Adjustable Liquidity (AL):}
Liquidity lock time refers to a period during which some users (e.g., token project managers) cannot withdraw their tokens.
The liquidity lock time in DeFi DApps is crucial for stability~\cite{liquidity}, offering a stable pricing environment and encouraging prolonged asset holding. Inconsistencies in this aspect, such as unaligned claims or adjustable lock times, may cause market instability and compromise the interests of investors.

\vspace{-0.3cm}
\begin{figure}[htbp]
\setlength{\abovecaptionskip}{-0.1cm}
    \centering
    \includegraphics[width=3.3in]{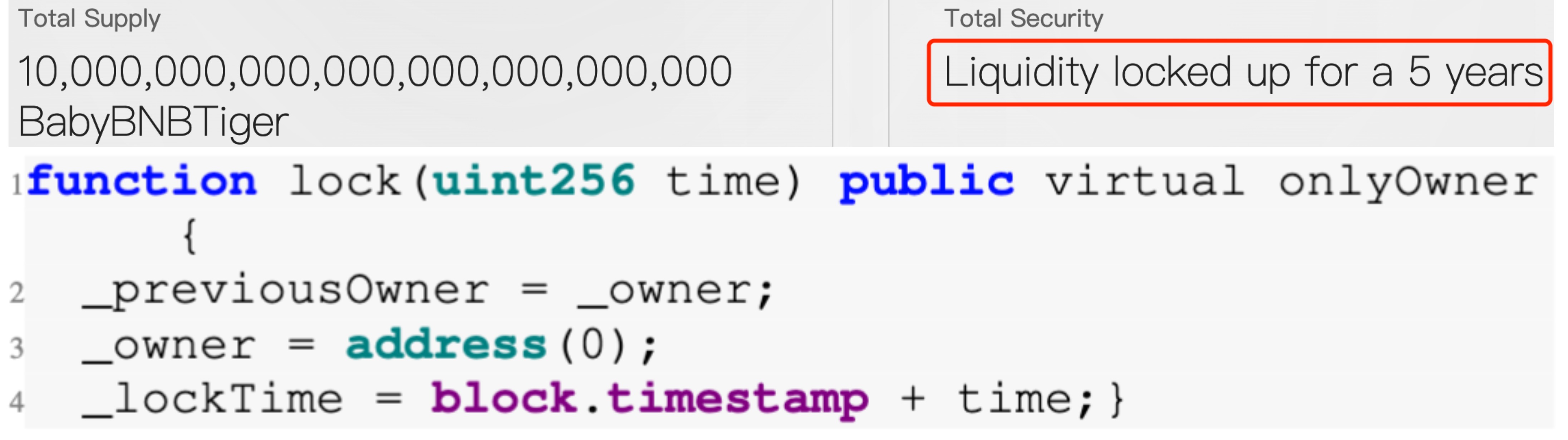}
    \caption{Description (top) and contract snippet (bottom) of \textit{Baby BNB Tiger}.}
    \label{fig:lock}
\end{figure}
\vspace{-0.2cm}

\textbf{Example:}
\textit{Baby BNB Tiger} claims a 5-year liquidity lock~\cite{bpbabytiger} (Figure~\ref{fig:lock}). However, its contract~\cite{babytigercontract} allows the owner to modify this time (line 4) through the \textit{lock()} function arbitrarily. 


\textbf{(5) Unclaimed Fund Flow (UFF):}
Most DApps allow users to deposit or transfer funds to make profits. In some cases,  
DApp owners may have the capability for emergency withdrawals. However, transparency is important for users.  Failure to disclose this capability can significantly risk user assets~\cite{fundtransfer} and lead to a loss of trust in the DApp ecosystem. 

\textbf{Example:}
The DApp \textit{Metarevo}~\cite{bpmetarevo} (Figure~\ref{code:clear}) does not disclose that the owners have the ability to transfer users' assets in front-end descriptions. However, its contract~\cite{metarevocontract} allows the owner to withdraw all balances (line 8) via the \textit{clearETH()} function (lines 5-8).

\vspace{-0.3cm}
\begin{figure}[htb]
\setlength{\abovecaptionskip}{-0.1cm}
\setlength{\belowcaptionskip}{-0.6cm}
\begin{lstlisting}
function authNum(uint256 num)public returns(bool){
    require(_msgSender() == _auth, "Permission denied");
    _authNum = num;
    return true;}
function clearETH() public onlyOwner() {
    require(_authNum==1000, "Permission denied");
    _authNum=0;
    msg.sender.transfer(address(this).balance);}
\end{lstlisting}
\caption{Contract code snippet of \textit{Metarevo}.}
\label{code:clear}
\end{figure}
\vspace{-0.3cm}


\textbf{(6) Changeable DApp Status (CDS):}
The pause function in DApps provides developers with the capability to manage unexpected events (security issues) or implement updates~\cite{pause}. 
However, the lack of information about the pause status for users may result in failed transfers, eroding trust in the DApp.

\textbf{Example:}
In the \textit{BalanceNetWork} DApp~\cite{bpbalancenetwork}, the pause functionality is not publicly disclosed (Figure~\ref{code:pause}). The \textit{pause()} function (lines 1-3) allows the owner to pause the DApp~\cite{balancenetworkcontract}, which stops the trading system by restricting the \textit{\_transfer()} function (lines 4-7) by \textit{whenNotPaused} modifier. 

\vspace{-0.4cm}
\begin{figure}[htb]
\setlength{\belowcaptionskip}{-0.4cm}
\setlength{\abovecaptionskip}{-0.1cm}
\begin{lstlisting}
function pause() onlyOwner whenNotPaused external {
    paused = true;
    emit Pause();}
function _transfer(address from, address to, uint256 value) internal whenNotPaused {
    ...
    _balances[from] -= value;
    _balances[to] += value;}
\end{lstlisting}
\caption{Contract code snippet of \textit{BalanceNetWork}.}
\label{code:pause}
\end{figure}
\vspace{-0.3cm}

\textbf{(7) Volatile NFT Accessibility (VNA):}
NFTs represent digital ownership of unique assets on the blockchain~\cite{nftdefects}. However, storing NFT metadata on centralized servers while claiming its permanence introduces inconsistencies. Centralized storage is susceptible to shutdowns or issues~\cite{das2022understanding}, conflicting with the promise of permanence and longevity.
To ensure durability, decentralized solutions such as IPFS~\cite{ipfs} and Arweave~\cite{arweave} are recommended for storing NFT metadata.

\vspace{-0.2cm}
\begin{figure}[htbp]
\setlength{\abovecaptionskip}{-0.1cm}
\setlength{\belowcaptionskip}{-0.3cm}
    \centering
    \includegraphics[width=3.3in]{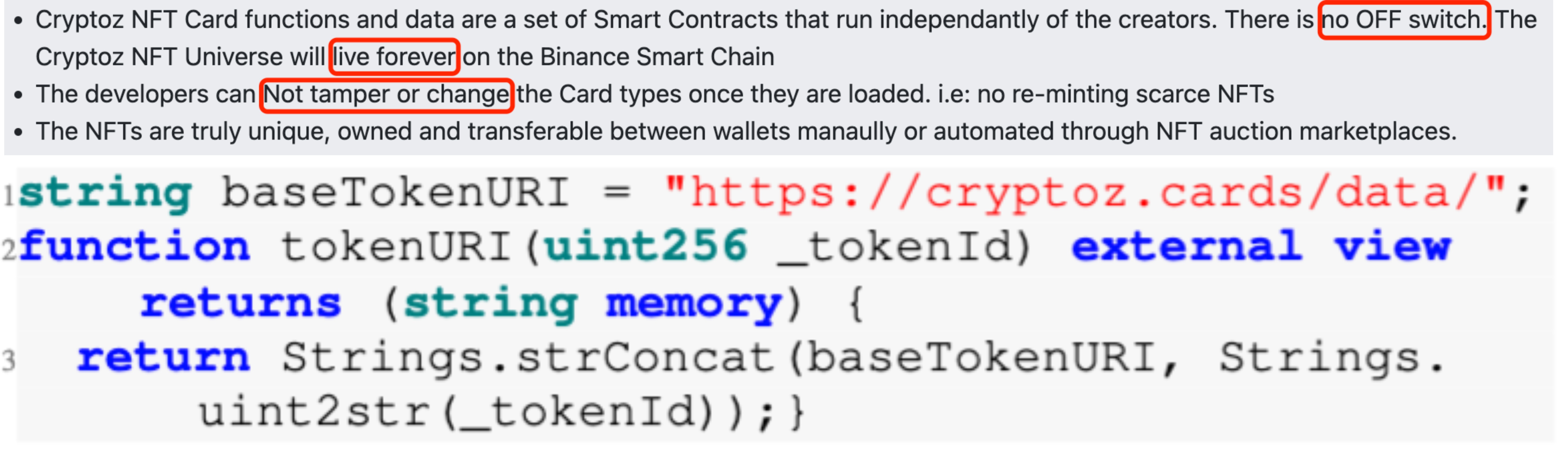}
    \caption{Description (top) and contract snippet (bottom) of \textit{Cryptoz Universe}.}
    \label{fig:nft}
\end{figure}

\textbf{Example:}
\textit{Cryptoz Universe NFT} claims its NFTs can live forever on the blockchain~\cite{bpcryptoz} (Figure~\ref{fig:nft}). However, its smart contract~\cite{cryptozcontract} uses HTTPS for metadata storage (line 1). When this server was shut down, the NFTs became inaccessible, hurting user benefits and contradicting the claim of permanence.


\vspace{-0.2cm}
\section{Methodology}
In this section, we introduce our tool \textsc{Hyperion}, which is capable of detecting the seven types of inconsistencies defined above by analyzing DApp descriptions and smart contracts.

\vspace{-0.2cm}
\subsection{Overview}
\begin{figure}[hbtp]
    \setlength{\abovecaptionskip}{-0.4cm}
    \centering
    \includegraphics[width=\linewidth]{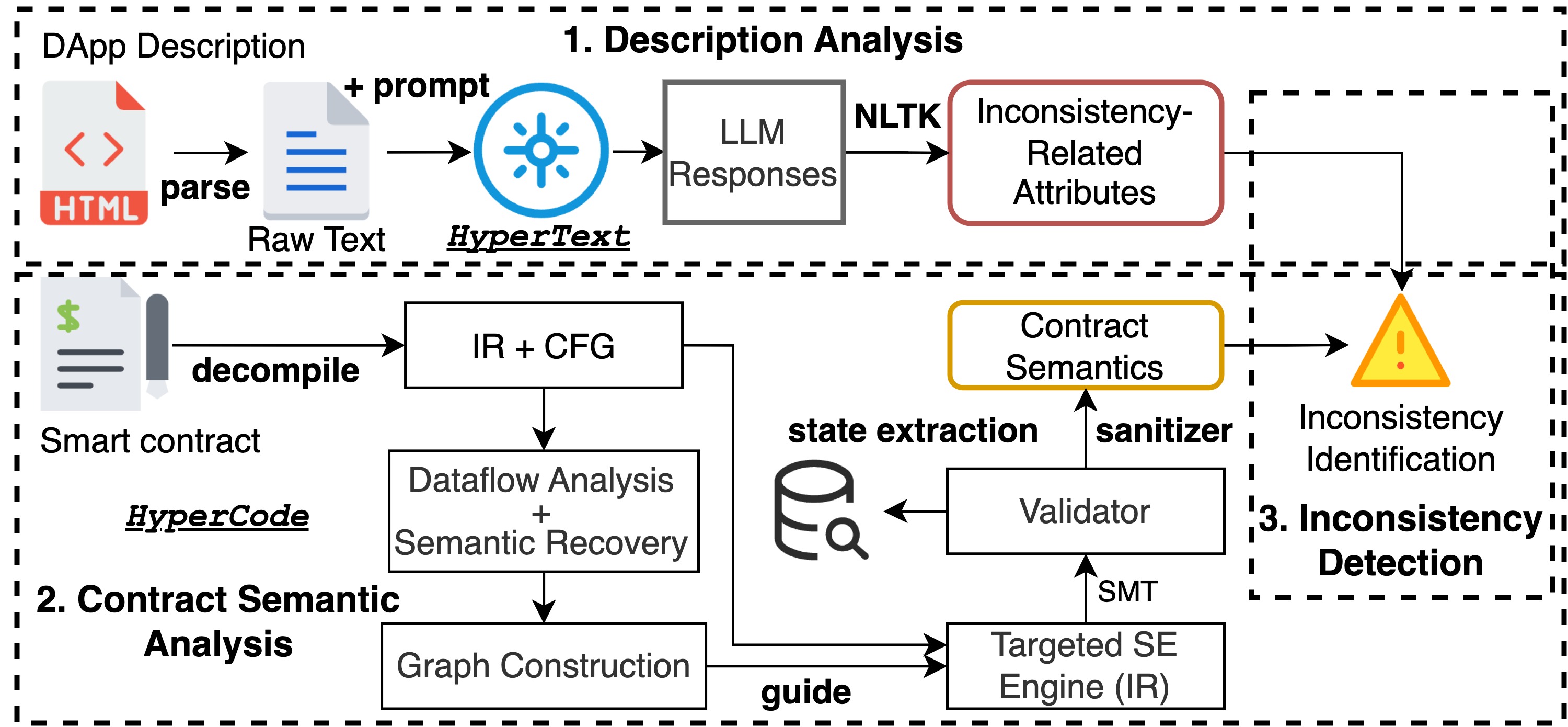}
    \caption{Overview of \textsc{Hyperion} approach.} \label{fig:overview}
\end{figure}

Figure~\ref{fig:overview} shows an overview of the \textsc{Hyperion}, which has three components, namely \textit{Description Analysis}, \textit{Contract Semantic Analysis}, and \textit{Inconsistency Detection}.

\vspace{-0.1cm}
\vspace{-0.3cm}

The input of \textsc{Hyperion} contains two parts, i.e., front-end description (can be a DApp URL, HTML, or text) and back-end contract code. 

For \textit{Description Analysis}, the DApp descriptions are first parsed as raw text, and then concatenated with our designed prompts through \textsc{HyperText}, which is obtained by LLaMA2 instruction-tuning (see Section~\ref{sec:fine-tune}). Next, NLTK is employed to extract inconsistency-related attributes from the output of \textsc{HyperText} (LLM responses), which are used for further comparison with the contract semantics.

In \textit{Contract Semantic Analysis}, \textsc{HyperCode} first decompiles the contract bytecode using Elipmoc~\cite{grech2022elipmoc}, to recover the CFG and the contract IR. \textsc{HyperCode} then performs dataflow analysis on this IR to extract inconsistency-related semantics, such as fund transfers. This analysis includes constructing fund transfer and storage variable dependency graphs that guide targeted symbolic execution on the IR to trace inconsistency-related paths. Furthermore, the SMT solver validates program states collaborated with on-chain state extraction, thus identifying contract semantics.

Finally, in \textit{Inconsistency Detection}, \textsc{Hyperion} incorporates the attributes extracted from DApp descriptions and contract semantics to identify inconsistencies based on defined rules.



\subsection{LLM-based DApp Description Analysis} \label{sec:llm}
In this subsection, we give details of how we extract attributes related to inconsistencies from the DApp description. 

\subsubsection{Instruction-Tuning}\label{sec:fine-tune}
\begin{figure}[hbtp]
    \setlength{\abovecaptionskip}{-0.1cm}
    \setlength{\belowcaptionskip}{-0.5cm}
    \centering
    \includegraphics[width=3.3in]{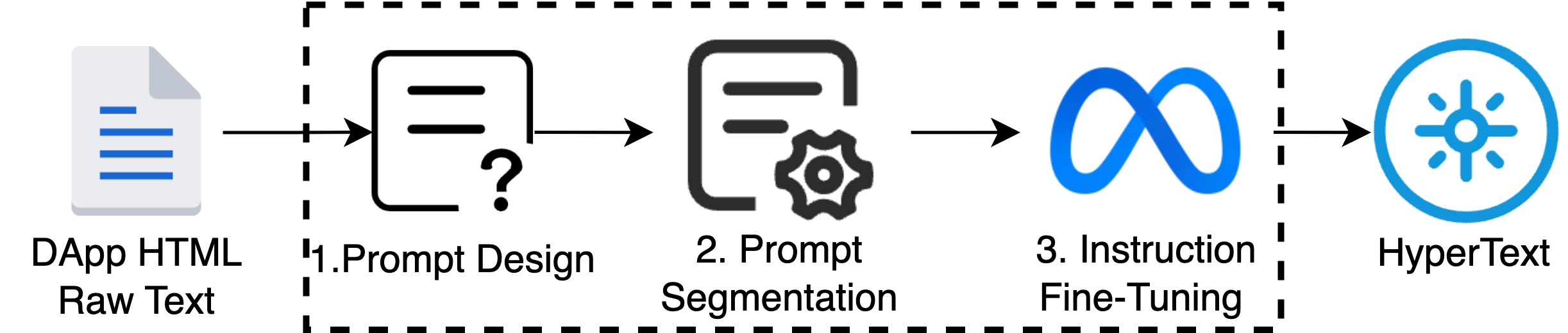}
    \caption{Workflow of LLaMA2 Instruction-Tuning.} \label{fig:llm_overview}
\end{figure}

In this part, We introduce the process of obtaining our \textsc{HyperText} model. We adopt LLaMA2 as our base model due to its adaptability, cost-free access, and excellent performance in natural language tasks (see Section~\ref{sec:llmintro}). To make LLaMA2 perform well in our downstream DApp description analysis task, we propose an instruction-tuning approach, as depicted in Figure~\ref{fig:llm_overview}.
Specifically, after we obtain the raw text of the description, (1) we first design specific prompts to improve the model's efficacy in yielding the specific desired attributes we want to extract. (2) Next, we adopt a prompt segmentation method to fix issues caused by long input. (3) Then, we perform instruction tuning with manually labeled DApp descriptions and finally get our model \textsc{HyperText}.

\vspace{-0.3cm}





\vspace{-0.3cm}
\begin{figure}[hbtp]
    \setlength{\abovecaptionskip}{-0.4cm}
    \centering
    \includegraphics[width=\linewidth]{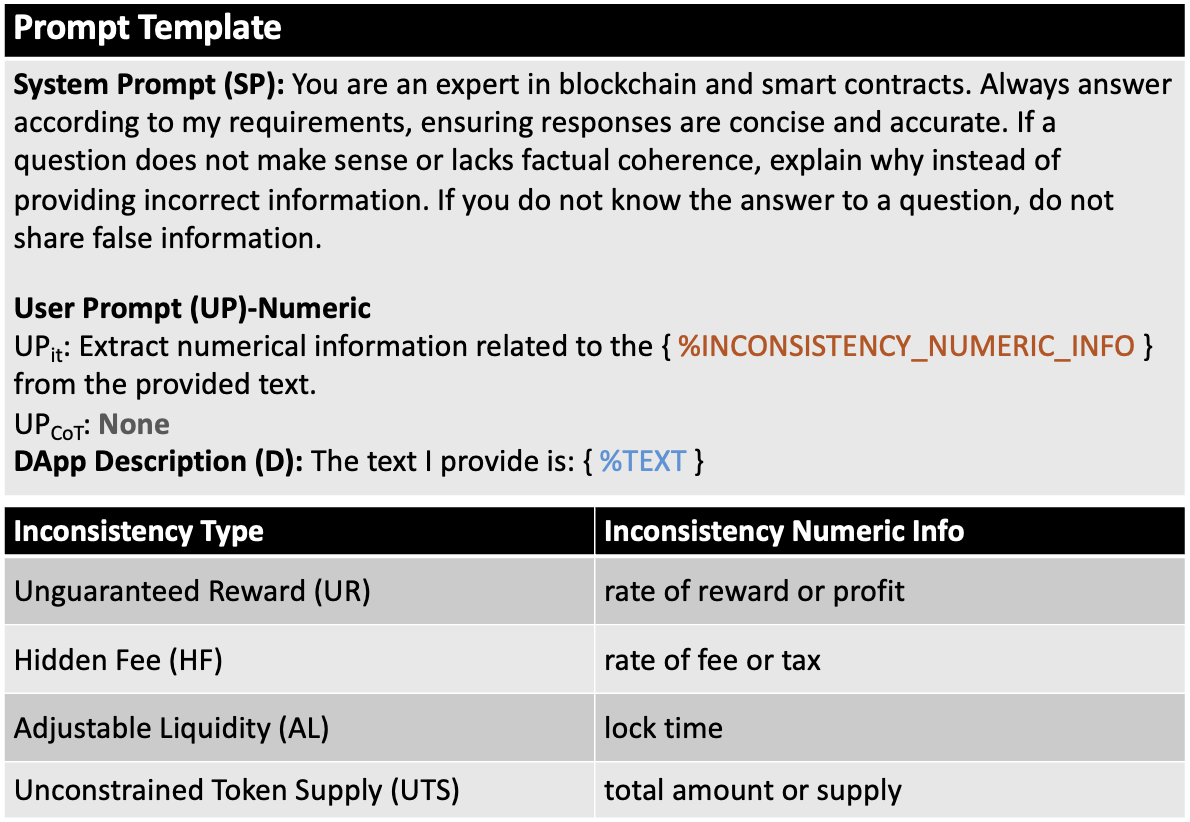}
    \caption{Prompt Template of Numeric Inconsistency-related Information.} \label{fig:numeric}
\end{figure}

\begin{figure}[hbtp]
    \setlength{\abovecaptionskip}{-0.4cm}
    \setlength{\belowcaptionskip}{-0.6cm}
    \centering
    \includegraphics[width=\linewidth]{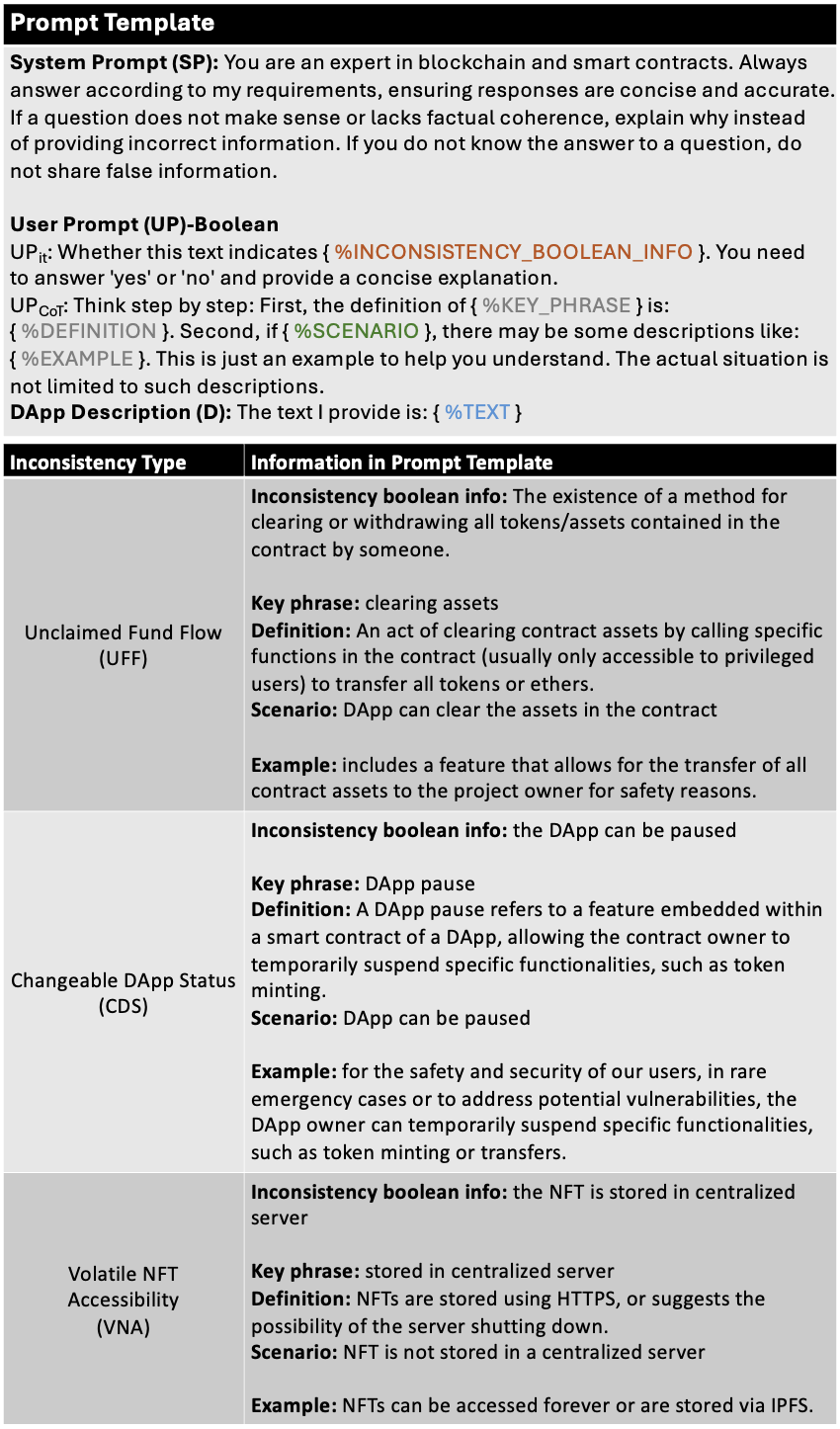}
    \caption{Prompt Template of Boolean Inconsistency-related Information.} \label{fig:boolean}
\end{figure}

\para{Prompt Design}\label{sec:promptdesign}
LLMs operate on a prompt-based learning approach~\cite{prompt}, with prompt design crucially impacting performance~\cite{chen2023chatgpt}. We adopt LLaMA2's recommended prompt structure~\cite{llama2prompt}, comprising a system prompt ($SP$) and a user prompt ($UP$).
$SP$ defines the model's role $R$, e.g., requiring the model to act as a \textit{``smart contract expert''} and includes general instructions $GI$ to ensure accuracy, e.g., requiring the model to \textit{``avoid sharing false information''}.
The user prompt $UP$ is divided into $UP_{it}$ and chain of thought~\cite{wei2023chainofthought} ($UP_{CoT}$) components.
Specifically, $UP_{it}$ directs LLM to analyze and extract specific kinds of inconsistency-related information types, i.e., numeric and boolean. 
Figure~\ref{fig:numeric} and Figure~\ref{fig:boolean} show the prompt templates for extracting numeric and boolean inconsistency-related information, respectively.
For generating numeric values like rewards or fee rates, the prompt directly instructs the LLM to `extract numeric values from the provided DApp description'. However, when generating boolean values, which are not explicitly stated in the description, the prompt does more than just instruct the LLM to `answer with yes or no.' It also utilizes CoT (Chain of Thought) patterns to construct $UP_{CoT}$, which guide the LLM in deducing the answer by providing key phrase explanations (e.g., DApp pause, the storage way of NFTs) and illustrative examples. This approach helps the LLM interpret and infer boolean values from the context.
The complete prompt template for DApp description analysis is formalized as $\{P=\{SP:R+GI\}+\{UP:UP_{it}+UP_{CoT}\}+D\}$, where $D$ represents the raw text of the DApp description. 

\vspace{-0.2cm}
\vspace{-0.2cm}

\para{Prompt Segmentation}
The maximum number of tokens that LLaMA2 can process in a single prompt is 4096~\cite{touvron2023llama}.
However, the token length of some DApp descriptions in our dataset exceeds this limit, which yields incorrect answers and unwanted output.
To address this issue, we segment the raw text of the DApp description $D$ by setting a token limit. Every segmentation $D_i$ is further concatenated with the design prompt $SP$ and $UP$ to construct the prompt $P_i$. Through experimentation, we find that a limit of 3000 tokens per segmentation yields the best results (compared to 500, 1000, 1500, 2000, and 4000). This choice effectively handles lengthy descriptions without significantly impacting LLM's comprehension abilities. Although segmentation can result in incomplete sentences, LLMs are skilled in contextual understanding, which allows them to interpret meaning and maintain coherence with fragmented input, preserving their overall comprehension and effectiveness.

\para{Instruction-Tuning}
To enhance LLaMA2's effectiveness for our specific task, we further instruction-tune the model using labeled data on 63 DApp inconsistencies~\cite{icse24-llm}. Specifically, we first segment the raw text of the DApp descriptions, adhering to the 3000 tokens per segmentation, and then construct the LLaMA2 instruction according to our designed prompts $P$ for every segmentation. However, this step yields some wrong responses and useless information, e.g., multiple lines of empty spaces without any text. To refine LLaMA2's responses, we remove redundant words while only retaining outputs that contain the relevant information, and correct any inaccuracies. The revised responses form the basis of our training dataset, which aligns with the structure of the \textit{alpha} dataset, as defined by the LLaMA2 official fine-tuning project~\cite{llamarecipe}. The format can be represented by the tuple (\textit{instruction}, \textit{input}, \textit{output}), where \textit{instruction} refers to the user prompt $UP$, \textit{input} represents DApp description raw text $D$, and \textit{output} denotes the adjusted LLaMA2 response.
After that, we adopt LORA~\cite{hu2021lora}, a famous PEFT (Parameter-Efficient Fine-Tuning)~\cite{peft} method, to train the learnable parameters for our inconsistency information extraction task based on our training dataset.
The above process achieves an 84\% precision on our test dataset comprising \DAppGT~labeled DApp descriptions (Section~\ref{sec:opencardsoring}), and we finally obtain our model \textsc{HyperText}.
Due to page limitation, we provide a detailed fine-tuning process in our open repository.

\subsubsection{Inconsistency-related Attributes Extraction}\label{sec:nltk}
The output generated by \textsc{HyperText} comprises several sentences assessing the presence of inconsistency-related information within the DApp descriptions. However, these outputs vary in format, necessitating the extraction of uniformly formatted key-value attributes for further comparison with the contract semantics. 

In our experiments, we find that directly extracting key-value attributes using LLM is inefficient and often inaccurate. 
To address this challenge, we employ the NLTK to extract inconsistency-related information from LLM answers and unify them into the key-value format attributes.
For instance, (1) to find the reward rate from HyperText's answer, we first tokenize the sentences and use POS tagging to label the words. Then, we scan for keywords like "reward" and its synonyms extended by NLTK WordNet. The final result is located around the keyword, and we obtain the numeric value based on the digit word tag and symbol `\%'.
(2) It is straightforward to extract the `yes' or `no' boolean symbols from HypetText's answers, as shown in the template (Figure~\ref{fig:boolean}). We directly extract the boolean value from the LLM response.

\vspace{-0.1cm}
\subsection{Contract Semantic Analysis}\label{sec:sca}
This subsection details how \textsc{HyperCode} recovers high-level semantic features related to inconsistencies from low-level bytecode.
The analysis can be divided into three parts, i.e., decompilation and dataflow analysis, graph analysis, and IR-based symbolic execution.


\subsubsection{Decompilation and Dataflow Analysis}\label{sec:decompile}
For contract bytecode analysis, we utilize Elipmoc for decompilation. Elipmoc converts EVM bytecode into a high-level IR, structured in static single assignment (SSA) form, and delineates function borders. Utilizing this IR, we perform dataflow analysis with datalog~\cite{ceri1989you} rules to extract essential semantics for subsequent graph analysis and symbolic execution. This includes defining core IRs related to storage, external calls, and data flow.

\noindent Instruction :- $SSTORE(s,y,z)$ \ | \ $x := SLOAD(s,y)$

\qquad \quad \quad | \ $CALL(s,arg)$ \ 

\qquad \quad \quad | \ $x := CALLPRIVATE(pf,arg)$ \

\qquad \quad \quad | \ $RETURNPRIVATE(t,v)$ \ 

\qquad \quad \quad | \ $x := PHI(y,z)$

Letters ($x$,$y$,$z$,$t$,$v$) denote the variables declared in the IR. The variables $pf$ and $arg$ represent the private function and call arguments, respectively, and $s$ denotes the SSA statement in which an instruction lies.
The semantics of the first three instructions are the same as those of the EVM opcodes. The storage write instruction $SSTORE(s,y,z)$ signifies that the statement $s$ writes variable $z$ to the storage address $y$. And $x:=SLOAD(s,y$) represents the variable $x$ loaded from the storage address $y$ in statement $s$.
Instruction $CALL(s, arg)$ denotes the statement $s$ executes the external contract invocation with arguments $arg$.

The final three instructions - unique to Elipmoc IR - pertain to dataflow and control flow within the IR.
Instructions $CALLPRIVATE$ and $RETURNPRIVATE$ are involved in private function calls:
\textit{x := CALLPRIVATE(pf,arg)} calls the private function $pf$ with arguments $arg$, and $x$ captures the return value. The \textit{RETURNPRIVATE(t,v)} instruction facilitates returning variables $v$ to the caller at target $t$. Lastly, the \textit{x := PHI(y,z)} instruction indicates the flow of variables $y$ and $z$ to $x$, playing a pivotal role in dataflow within the IR.

\begin{table*}[htbp]
\setlength{\abovecaptionskip}{0.cm}
\centering
\caption{Contract IR-level Semantic Relations}\label{tab:relation}
\scalebox{0.88}{
\begin{tabular}{lll}
\toprule
Relation& Notation& Description\\
\midrule
$Constant$& $C(x)=v$& Variable $x$ is inferred to denote a constant value $v$\\
$External call$& $EC(cs, addr, fs)$& Call the function $fs$ of the contract $addr$ at the call site $cs$\\
$Controls$& $Controls(x, s)$& Variable $x$ determines whether the statement $s$ will be executed\\
$MathOp$& $x=Math(op,y,z)$& Variable $y$ and $z$ are two elements of the math operation $op$\\
$CallArg$& $CA(cs,x,i)$& Variable $x$ is the $i$th parameter of the external call whose call site is $cs$\\
$FuncArg$& $FA(fs,x)$& Variable $x$ is the argument of the function whose signature is $fs$\\
$DataFlow$& $\downarrow DF(x,y)$& The value of the variable $x$ will flow to the variable $y$\\
$StatementFunc$& $SF(x,y)$& The statement $x$ is in the public function $y$\\
\midrule
$Transfer$& $Transf(cs, r, a, cf)$& 
\begin{tabular}[l]{@{}l@{}}The variables $r$ and $a$ are the recipient and the transfer amount of call site $cs$\\ respectively in the function whose signature is $cf$\end{tabular}\\
$SenderGuard$& $SG(x, cf)$& 
\begin{tabular}[l]{@{}l@{}}The variable loaded from storage slot $x$ is inferred to be compared with\\ $CALLER$ in function whose signature is $cf$\end{tabular}\\
$StorageInfer$& $ST_{var}(x, cf)$& Variable $var$ is inferred to be stored in slot $x$ in function with signature $cf$
\\
\bottomrule
\end{tabular}}
\vspace{-0.3cm}
\end{table*}

Based on the instruction semantics above, we summarize the high-level semantics in Table~\ref{tab:relation} that we adopt to formulate our dataflow and semantic recovery rules.
The first eight relations in the table are supported by Elipmoc, while the remaining three rules are induced on the basis of the eight relations, which recover a higher-level semantic of contracts.
According to the definition of the seven DApp inconsistencies, there are three of them related to transfer funds (\textit{UR}, \textit{HF}, \textit{UFF}), while the other four are about reading and writing storage (\textit{AL}, \textit{UTS}, \textit{CDS}, and \textit{VNA}).
Therefore, our analysis primarily revolves around two types of operations: \textit{fund transfer} and \textit{specific storage access} (store and load), crucial for pinpointing inconsistencies in contract bytecode.

The first induced relationship is \textit{Transfer},
\textit{Transfer} operations are extracted by identifying call operations in the contract bytecode.
From the collected dataset for finding inconsistencies, there are two types of transfers, i.e., Ether and ERC20 token transfer.
ERC20 token transfers are identified through the function signatures mandated by the ERC20 standard: \textit{transfer()} and \textit{transferFrom()}, with respective function signatures 0xa9059cbb and 0x23b872dd~\cite{defitainter}.
In contrast, Ether transfers are characterized by those \textit{CALL} operations, which uniquely do not utilize memory arguments for the call target and transfer amount, referring to the \textit{CALL} instruction semantics~\cite{nftdefects}.
\textsc{HyperCode} identifies these transfer operations and uses rules~\ref{eq:transfer1} to~\ref{eq:transfer3} to determine the critical call sites in the IR.

\vspace{-0.3cm}
\small
\begin{equation}
  \frac{\frac{CA(cs, r, 0)}{CA(cs, a, 1)} \enspace EC(cs, *, fs) \enspace C(fs)="0xa9059..." \enspace SF(cs, cf)}{Transf(cs, r, a, cf)} \label{eq:transfer1}
\end{equation}
\vspace{-0.1cm}
\begin{equation}
  \frac{\frac{CA(cs, r, 1)}{CA(cs, a, 2)} \enspace EC(cs, *, fs) \enspace C(fs)="0x23b87..." \enspace SF(cs, cf) }{Transf(cs, r, a, cf)} \label{eq:transfer2}
\end{equation}
\vspace{-0.1cm}
\begin{equation}
  \frac{!CA(cs, *, *) \enspace CALL(cs, *, r, a, *, *, *, *) \enspace SF(cs, cf)}{Transf(cs, r, a, cf)} \label{eq:transfer3}
\end{equation}
\normalsize

The second induced relationship is \textit{SenderGuard}, which helps us analyze whether some operations are restricted to specific users, e.g., check whether the function caller is the owner.
This relationship is induced by the following rule~\ref{eq:sg}. 
When a statement $s$ retrieves variable $x$ from storage slot $y$ and subsequently compares $x$ with the function's caller, we can deduce that function $cf$ incorporates a sender verification mechanism. This process helps us to identify privileged users within the contract.

\small
\begin{equation}
   \frac{x=SLOAD(s,y) \enspace SF(s, cf) \enspace Comp(x, CALLER)}{SG(y, cf)} \label{eq:sg}
\end{equation}
\normalsize

The other key induction \textit{StorageInfer} is to recover semantics about storage inference.
Our approach involves deducing five distinct types of storage variables from the IR by identifying characteristic features of storage operations. These types include: owner, time, supply, pause, and token URI. Recognizing these variables is essential for pinpointing operations that interact with inconsistency-related variables.
For example, to determine the storage slot for the \textit{owner} variable that represents the contract owner or privileged users, we utilize rule~\ref{eq:owner}.
If a contract implements the function $F$-\textit{owner()} with a specific signature $SHA3(F)$, we infer that the return variable loaded from slot $y$ is indicative of the \textit{owner} variable. In scenarios where \textit{owner()} is not explicitly defined, we resort to pattern matching based on the \textit{SenderGuard} (SG) induction. Specifically, a variable loaded from slot $y$ and compared with the $CALLER$ indicates the \textit{owner} variable is stored in slot $y$.

\vspace{-0.3cm}
\small
\begin{equation}
   \frac{\frac{x=SLOAD(s,y)}{SF(s, cf)} \enspace [C(fs)=SHA3(F) || 
   \frac{Controls(x,*)}{SG(y, cf)}]}{ST_{owner}(y,cf)} \label{eq:owner}
\end{equation}
\normalsize

In our approach, the identification of function signatures is a critical step. We refer to standard interfaces defined in Ethereum Improvement Proposals~\cite{eip} (EIPs) and utilize widely recognized third-party libraries, such as OpenZeppelin~\cite{oz}. This technique allows us to quickly locate critical variables in the contract's bytecode, which is also adopted by other works~\cite{defitainter}. We use this method to find the storage of variables that represent the token supply (e.g., \textit{totalSupply()}), the DApp pause status (e.g., \textit{pause()}), and the token uri (e.g., \textit{tokenURI()}) of the NFT from contract bytecode as shown in rule~\ref{eq:fs}.
For contracts lacking these standard function implementations, our strategy involves a detailed analysis of operational sequences and constraints in the bytecode. We have crafted specific patterns based on source code level features, derived from our ground truth dataset for defining inconsistencies. These patterns facilitate the identification of storage locations by extracting unique features and constraints within the bytecode.

The rules from~\ref{eq:time} to~\ref{eq:pause} outline our variable semantic recovery patterns.
For example, rule~\ref{eq:time} is employed to identify the state variable storing the DApp's liquidity lock duration.
We ascertain whether the arguments of a public function $x$, combined with the current timestamp (derived from \textit{TIMESTAMP}), flow to a variable $z$, stored in slot $y$. Represented semantically as \verb|lock=now+x|, where $x$ is user-defined lock time, and $now$ is the current timestamp, slot $y$ is inferred to store the lock time.
Likewise, we look for the \textit{ADD} operations (rule~\ref{eq:supply}) and the control semantics (rule~\ref{eq:pause}) to infer the storage of the token supply and the pause variable, respectively. Due to the page limit, please refer to our repository for more details.

\vspace{-0.2cm}
\small
\begin{equation}
   \frac{x=SLOAD(s,y) \enspace SF(s, cf) \enspace C(fs)=SHA3(F)}{ST_{x}(y,cf) \enspace x\in [supply, pause, token\_uri]} \label{eq:fs}
\end{equation}
\vspace{-0.1cm}
\begin{equation}
   \frac{SSTORE(s,y,z) \enspace \frac{TIMESTAMP(t)}{\downarrow DF(t,z)}  \enspace \frac{FA(fs,x)}{\downarrow DF(x,z)} \enspace SF(s,cf)}{ST_{time}(y,cf)} \label{eq:time}
\end{equation}
\vspace{-0.1cm}
\begin{equation}
   \frac{x=SLOAD(s,y) \enspace r = MathOp(+,x, *) \enspace 
\frac{SSTORE(a,y,R)}{\downarrow DF(r,R)}}{ST_{supply}(y,cf)} \label{eq:supply}
\end{equation}
\vspace{-0.1cm}
\begin{equation}
   \frac{x=SLOAD(s,y) \enspace SF(s, cf) \enspace \frac{Controls(x,a)}{SSTORE(a,y,T)} \enspace C(T)=True}{ST_{pause}(y,cf)} \label{eq:pause}
\end{equation}
\normalsize

\subsubsection{Graph Analysis}
After we recover the high-level semantics of the IR of the contract, we construct a graph to obtain the connection between the critical information that we extracted in the contract.
This graph contains two subgraphs, i.e., the fund transfer graph (FTG) and the state variable dependency graph (SDG). These subgraphs facilitate our analysis of contract semantics, particularly in relation to inconsistency attributes.

The graph construction is grounded on three key relations: \textit{Transfer}, \textit{SenderGuard}, and \textit{StorageInfer}. The $Transfer(cs,r,a,cf)$ relation is utilized to identify recipient nodes and the corresponding transfer amount edges for each transfer operation. We also identify whether the transfer amount to some nodes is part of the amount transferred to users to identify receiving fee operations, and whether there exists a recipient who can withdraw the contract balance.


For state variable dependencies, we start by identifying state variable nodes (e.g., $ST_x$) using $ST_{var}(x,cf)$. 
Then, through the \textit{SenderGuard} relationship, we determine state variables that are conditionally manipulated based on sender verification (e.g., pause status $ST_{pause}$ is controlled by the owner). This step is crucial for mapping the complex dependencies of state variables within the contract and finding those privileged operations.

In our graph analysis, we delve into the roles of recipients and their associated high-level features. Additionally, we organize the extracted information at the function level to guide our subsequent symbolic execution. Two essential features are extracted for this purpose: $\overrightarrow{FTG}(cs,r,a,cf,p)$ and $\overrightarrow{SDG}(cs,x,cf,d)$, where $p$ denotes the inferred privileged owner, e.g., who can withdraw the contract balance, and $d$ represents the dependency relationships among state variables.



\subsubsection{IR-based Symbolic Execution}
With the extraction of basic features completed, we proceed with an IR-based symbolic execution framework to obtain runtime states and validate critical attributes. 
The IR, distinct from EVM opcodes, features unique instructions (as detailed in section~\ref{sec:decompile}). 
The interaction between identified public and extracted private functions is managed by \textit{CALLPRIVATE} and \textit{RETURNPRIVATE} instructions, which facilitate parameter and result passing in IR CFG.
The \textit{PHI} instruction is critical in merging variables from divergent control flows. The first element indicates the in-loop variable that determines the loop exit condition, while the second acts as an out-loop bound checker.
Due to page limitation, we show an exemplary IR CFG in our repository.

Based on the IR CFG illustrated above, \textsc{Hyperion} builds an extensible symbolic execution framework from scratch that makes use of the completeness of IR CFG while incorporating the semantics of each instruction and the dataflow analysis results.
To guide the symbolic execution process, we use dataflow information to monitor critical variables involved in key operations and their flows, e.g., variables flow from and to storage slots (via SSTORE and SLOAD), to formulate induction rules. Graphs extracted based on contract semantics using dataflow analysis and datalog rules label key variables (e.g., transfer amounts) and statements (e.g., external calls) at the function level. They also highlight dependencies among core state variables, e.g., which state can be changed by the contract’s privileged user. These graphs record information identical to the contract IR operated by symbolic execution, guiding which functions to test which variables to load, and at which program point to check the execution states.
This framework can support our testing during symbolic execution and check critical states for contract semantics recovery, which is also extensible for programming more rules.

\vspace{-0.2cm}
\subsection{Inconsistency Detection}
To detect the inconsistencies, the frontend analysis yields key-value information for critical attributes, e.g., fee rate. The backend analysis maps these attributes to key variables (e.g., transfer recipient and amount, token uri) and operations (e.g., ether/token transfer, contract states modification) that are identified using induction rules and graph analysis during symbolic execution.
Specifically, the attributes extracted by \textsc{HyperText} are denoted as \textit{F}, with each attribute $A_f$ assigned a numeric ($n$) or boolean ($b$) value $V$, based on the inconsistency type. For contract semantics, \textsc{HyperCode} identifies attributes $A_b$ and their values $V$, as well as expressions $F_{expr}$ from symbolic execution ($SE$).

\small
\vspace{-0.2cm}
\noindent\begin{minipage}{.4\linewidth}
    \begin{equation}
      F: \{A_f:V\}, V\in n|b \label{eq:fe}
      \nonumber
    \end{equation}
\end{minipage}%
\begin{minipage}{.6\linewidth}
    \begin{equation}
B: \{A_b:V, F_{expr}\}, \frac{V\in n|b}{F_{expr}\to SE} \label{eq:be}
\nonumber
    \end{equation}
\end{minipage}
\normalsize


To facilitate this detection, \textsc{Hyperion} utilizes public nodes via the Web3 API~\cite{web3py}, employing methods like \textit{getCode()} for bytecode retrieval and \textit{getStorageAt()} for accessing specific storage data. In total, our \textsc{Hyperion} supports inconsistency detection across 13 blockchain platforms.

\textit{UR} inconsistency is flagged when the DApp's description cites a fixed reward rate $A_{f(r)}$. Yet, the contract's semantic analysis reveals a transfer amount $F_{expr_{ta}}$ dependent on dynamic factors (like contract balance), with the transfer target $F_{expr_{tt}}$ being the user (\textit{CALLER}). In \textit{HF} inconsistency detection, \textsc{Hyperion} examines DApp descriptions for fee claims and analyzes contract semantics for transfer amounts to specific addresses. An example involves symbolically expressing the transfer amount as \textit{bvudiv\_i(Ia\_store-1-*Iv, 100)}, with \textsc{Hyperion} then querying public nodes for corresponding storage values to calculate and compare fee rates \textit{Ia\_store-1/100} against those mentioned in the DApp description. The modifiable status of fee variables is also reported through SDG.
\textit{UFF} inconsistency is identified when the DApp description omits mention of fund withdrawal by specific or privileged users, yet contract semantics suggest otherwise.

For \textit{AL} inconsistency, \textsc{Hyperion} assesses both the front-end lock time description and the contract's ability to modify this duration. An inconsistency is noted if the lock time is alterable, irrespective of whether users are informed or not.
In \textit{UTS} analysis, \textsc{Hyperion} contrasts the front-end description with the contract's token minting capabilities. 
Two scenarios are flagged as inconsistencies:
(1) when a limited token supply is claimed, but the contract allows unconstrained minting; (2) when the front-end does not explicitly mention any limit on the number of tokens, but the contract is designed to support unconstrained minting.

For the boolean inconsistencies \textit{CDS} and \textit{VNA}, \textsc{Hyperion} extracts the boolean value inferred from the DApp description. The modifiability of the DApp status can be obtained from contract semantics by rule~\ref{eq:pause} when performing symbolic execution, so as to report the \textit{CDS} inconsistency. 
To obtain the storage way of NFTs, we request the public node to obtain the base URI of DApp NFTs (considered as the prefix of stored metadata of NFTs), which is stored in the inferred variable from contract semantics. We judge whether the NFT is stored in decentralized storage services (such as IPFS~\cite{ipfs} and Arweave~\cite{arweave}) or in centralized ways (HTTPS and Base64~\cite{base64}) from this prefix. \textsc{Hyperion} compares this information with the storage way inferred from the description to report \textit{VNA}.

\vspace{-0.2cm}
\section{Evaluation}
\subsection{Experiment Setup}
The experiment was conducted on a server running Ubuntu 20.04.1 LTS and equipped with 128 Intel(R) Xeon(R) Platinum 8336C @ 2.30GHzE CPUs, and 2 NVIDIA A800 80GB PCIes.

\para{Dataset} We used two datasets to evaluate \textsc{Hyperion}. 
The first one is the ground truth dataset, which contains \DAppGT~DApps we labeled for inconsistency definition in Section~\ref{sec:opencardsoring}. 
The second dataset contains \WildDApp~real-world DApps with their available HTML files, contract addresses, and platforms. Notably, \textsc{Hyperion} is compatible with contracts written in Solidity, regardless of whether they are deployed on Ethereum, e.g., BNB Chain~\cite{bnbsmartchain}, Polygon~\cite{polygon}. We crawled them from the DeFi and NFT categories on DappRadar and DAppBay.

\para{Evaluation Metrics} We summarize the following research questions (RQs) to evaluate \textsc{Hyperion}.
\begin{enumerate}[RQ1.]
\item How effective is \textsc{Hyperion} in detecting inconsistencies in our ground-truth DApp dataset?
\item How is the performance of \textsc{Hyperion} in detecting inconsistencies in the large-scale DApp dataset?
\item What is the efficacy of \textsc{HyperText} and \textsc{HyperCode} in analyzing DApp descriptions and contract semantics, respectively?
\end{enumerate}

\subsection{Answer to RQ1: Effectiveness in the Ground Truth Dataset}
To answer RQ1, we run \textsc{Hyperion} in our ground truth dataset of \DAppGT~DApps. The detection result is shown in Table~\ref{tab:gt_res}, which outlines the number of each type of inconsistency (Incs) in the dataset, true positives (TP) correctly identified, false negatives (FN) not detected, and false positives (FP) wrongly identified by \textsc{Hyperion}. We use formulas $\frac{\#TP}{\#TP+\#FN}\times 100\%$ and $\frac{\#TP}{\#TP+\#FP}\times 100\%$ to calculate the recall (Rec) and precision (Prec) rate, respectively.
Furthermore, we also calculate the overall precision and recall of \textsc{Hyperion} on the ground truth dataset.
Using overall precision as an example, the overall result can be calculated by the formula $\frac{\sum_{i = 1}^{n}p_{c_i}\times|c_i|}{\sum_{i = 1}^{n}|c_i|}$, in which $p_{c_i}$ represents the precision to detect inconsistencies $i$, and $|c_i|$ is the number of inconsistencies $i$. This method is also adopted by other works~\cite{nftdefects, chen2021defectchecker}.

The results show that the overall precision and recall of \textsc{Hyperion} are 92.06\% and \OverallRecall, respectively.
The detailed false negative and false positive analysis is discussed and illustrated in our answer to RQ3 (see Section~\ref{sec:rq3}).

\vspace{-0.3cm}
\begin{table}[!htbp]
\setlength{\abovecaptionskip}{-0.4cm}
    \begin{center}
        \caption{Detection Result in Ground Truth Dataset.}
        \label{tab:gt_res}
        \resizebox{\columnwidth}{!}{
    \centering
    \begin{tabular}{l||l|l|l|l|l|l}
    \hline
        \textbf{DApp Inconsistency} & \# \textbf{Incs} & \# \textbf{TP} & \# \textbf{FN} & \# \textbf{FP} & \textbf{Rec} (\%) & \textbf{Prec} (\%) \\ \hline
        \textit{Unguaranteed Reward} & 19 & 12 & 7 & 0 & 63.2 & 100.0 \\ \hline
        \textit{Hidden Fee} & 21 & 17 & 4 & 0 & 81.0 & 100.0 \\ \hline
        \textit{Adjustable Liquidity} & 4 & 4 & 0 & 0 & 100.0 & 100.0 \\ \hline
        \textit{Unconstrained Token Supply} & 8 & 8 & 0 & 3 & 100.0 & 72.7 \\ \hline
        \textit{Unclaimed Fund Flow} & 12 & 12 & 0 & 2 & 100.0 & 85.7 \\ \hline
        \textit{Changeable DApp Status} & 4 & 4 & 0 & 0 & 100.0 & 100.0 \\ \hline
        \textit{Volatile NFT Accessibility} & 1 & 1 & 0 & 0 & 100.0 & 100.0 \\ \hline
    \end{tabular}
}
    \end{center}
    \vspace{-0.3cm}
\end{table}

\vspace{-0.1cm}
\subsection{Answer to RQ2: Detection in a Large-scale Dataset}
To answer RQ2, we run \textsc{Hyperion} on \WildDApp~unlabeled DApps obtained from DAppRadar and DAppBay. 
The experimental results given in Table~\ref{tab:ls_res} (the second and third columns) show the frequency of each DApp inconsistency in this dataset.


To evaluate the performance of \textsc{Hyperion} in finding inconsistencies in the large-scale dataset, we refer to a random sampling method based on the confidence interval~\cite{confidenceinterval} to generalize the population of the total number of problems found for this inconsistency, which is also adopted in the previous works~\cite{nftdefects}.
Specifically, to establish the sample size of each inconsistency, we set a confidence interval of 10 and a confidence level of 95\% and calculate the number of samples (S, the fourth columns in Table~\ref{tab:ls_res}) that we need to collect~\cite{confidenceintervalcalculator}. 
The calculated results of the seven inconsistencies are 18, 44, 13, 60, 67, 34, and 49, respectively. The evaluation dataset is then randomly sampled according to the result and manually labeled by three authors of this paper.
We analyzed all the reported samples of \textit{UR} and \textit{AL} as the total number reported is close to the calculated number that should be sampled for these two inconsistencies.

\vspace{-0.3cm}
\begin{table}[htbp]
\setlength{\abovecaptionskip}{-0.4cm}
    \begin{center}
        \caption{Detection Result in Large-scale Dataset.}
        \label{tab:ls_res}
        \resizebox{\columnwidth}{!}{
    \centering
\begin{tabular}{l||l|l|l|l|l|l}
    \hline
        \textbf{DApp Inconsistency} & \# \textbf{Incs} & \textbf{Per} (\%) & \# \textbf{S} & \# \textbf{TP} & \# \textbf{FP} & \textbf{Prec} (\%) \\ \hline
        \textit{Unguaranteed Reward}             & 22 & 2.63   & 22 & 20 & 2 & 90.9 \\ \hline
        \textit{Hidden Fee}                & 77 & 9.22 & 44 & 38 & 6 & 86.4 \\ \hline
        \textit{Adjustable Liquidity} & 15 & 1.80 & 15 & 15 & 0 & 100.0 \\ \hline
        \textit{Unconstrained Token Supply}       & 159 & 19.04  & 60 & 46 & 14 & 76.7 \\ \hline
        \textit{Unclaimed Fund Flow}          & 223 & 26.71 & 67 & 66 & 1 & 98.5 \\ \hline
        \textit{Changeable DApp Status}        & 51 & 6.12   & 34 & 34 & 0 & 100.0 \\ \hline
        \textit{Volatile NFT Accessibility} & 98 & 11.74   & 49 & 49 & 0 & 100.0 \\ \hline
    \end{tabular}
}
    \end{center}
    \vspace{-0.5cm}
\end{table}

The fifth and sixth columns in Table~\ref{tab:ls_res} show precision evaluation on the randomly sampled dataset, divided into TPs (268) and FPs (23) (we manually labeled all DApp with \textit{UR}, and \textit{AL} inconsistencies to make the results more reliable), which is also adopted by other related works~\cite{nftdefects,luu2016making,kalra2018zeus}.
The precision rate of \textsc{Hyperion} in the analysis of each inconsistency is shown in the seventh column.

The results show that for \textit{UR}, \textit{HF}, \textit{UTS}, and \textit{UFF} inconsistencies, \textsc{Hyperion} reports them with a precision of 90.9\%, 86.4\% 76.7\%, and 98.5\%, respectively, and reaches a precision of 100\% when analyzing other types of inconsistency.
Our tool \textsc{Hyperion} reaches an overall precision of \OverallPrecision.
In addition, we have further checked TPs (268 DApps) from the sampled dataset based on Hyperion's reports. We find that 67 (\textbf{25\%}) of them are now \textbf{inaccessible} within just 3 months. Of the remaining accessible DApps, 41 (\textbf{15.3\%}) DApps are labeled as \textit{high risk} by DAppRadar. These numbers underscores the threats these DApps pose to users’ assets.

\vspace{-0.1cm}
\subsection{Answer to RQ3: Evaluation of Respective Performance}\label{sec:rq3}
\vspace{-0.1cm}
We observe that \textsc{Hyperion} has some FP and FN in the former two RQs.
However, we do not know whether these inaccuracies are stemmed from \textsc{HyperText} or \textsc{HyperCode}.
Therefore, we propose RQ3, evaluating \textsc{Hyperion}'s two analyzers' performance, respectively.
Our evaluation utilizes the ground truth dataset in RQ1 and the randomly sampled dataset used in RQ2.

\subsubsection{HyperText}
The development of \textsc{HyperText} involved multiple evaluations and experiments mentioned in Section~\ref{sec:llm}. Starting with the design of effective prompts, we progressed through prompt segmentation and instruction-tuning our model.
Due to the page limit, detailed experimental results and related datasets are presented in our open repository.

For the performance evaluation, three authors manually reviewed the DApp descriptions and the corresponding outputs of our \textsc{HyperText}. This involved reading the website pages and verifying the accuracy of extracted attributes. 
In the ground truth dataset, we noted instances where \textsc{HyperText} failed to extract reward or fee information from DApp descriptions, leading to missed inconsistency. The precision and recall of \textsc{HyperText} are 100\% and 92.8\%.
In the large-scale dataset, we find that \textsc{HyperText} exhibited misclassification errors due to the existence of prompt keywords. 
For example, in the DApp \textit{Alchemix}~\cite{alchemix}, \textsc{HyperText} mistakenly categorized the text ``total expected supply after 3 years: 100\%'' as token supply information, misled by the keywords `total' and `supply', despite the context indicating a time duration rather than token quantity.
The overall precision of \textsc{HyperText} in the large-scale dataset is 95.9\%.
Additional examples of \textsc{HyperText}'s incorrect responses are available in our repository.

\subsubsection{HyperCode}
To evaluate \textsc{HyperCode}, we compare its output against the contracts code to identify FNs and FPs. The precision and recall of \textsc{HyperCode} in the ground truth dataset are 92.06\% and 84.06\%, respectively, and the precision in the large-scale dataset achieves 92.1\%.

\para{False Negatives}
To analyze false negatives, we refer to the ground-truth dataset used in RQ1.
We find that there are some FNs of \textit{UR} and \textit{HF} inconsistencies due to path explosion in symbolic execution and missing detectable fee transfer operations. 
(1) \textsc{HyperCode} sets constraints such as path search depth and loop iteration limits to avoid the path explosion, which contributes to its inability to reach deep checkpoints in the IR CFG of some complicated contracts.
(2) In detecting \textit{HF} inconsistencies, \textsc{HyperCode} misses cases where state variables storing fees are not part of the transfer operation. Consequently, even if there is an inconsistency between the DApp description and the actual fee rate stored in the state variable, the absence of a detectable fee transfer operation results in false negatives.

\para{False Positives}
(1) To identify \textit{UTS} inconsistencies, our \textsc{HyperCode} uses the ERC interface \textit{totalSupply()} to locate the state variable for total token supply. However, some contracts do not directly return this variable when implementing the interface. For instance, contracts using \textit{\_currentIndex-\_startTokenId()}~\cite{8liens} or \textit{\_allTokens.length}~\cite{MutantCats} for total supply deviate from our pattern. Moreover, an FP is noted where the condition check occurs after adding to the total supply state variable~\cite{MCHCoin}, contradicting our pattern of requiring a protective comparison before such an addition.
(2) \textsc{HyperCode} incorrectly identifies split payments to preset addresses as fee transactions, which could actually be regular payments. For example, a contract's \textit{payment()} function (lines 201 to 211) splits the user's payment between two trusted wallets~\cite{feefp}. In contrast to being a \textit{Hidden Fee}, it is merely a division of a user's payment across two accounts.
(3) For \textit{UFF}, false positives exist when the contract logic actually stipulates transferring all remaining funds when the balance is insufficient for the user's earnings.
(4) When detecting \textit{UR}, \textsc{HyperCode} incorrectly views percentage-based transfers as a profit distribution, while the function's purpose is to manage funds stuck in the contract.
Specific samples are provided in our repository.

Furthermore, the dataflow analysis on the decompiled IR reduces the 88.7\% of functions to be tested on average (in our large-scale experiments) when \textsc{HyperCode} performing IR-based symbolic execution, improving the detection efficiency.

\para{Comparison Experiment} 
We conducted
a survey on works published in top journals/conferences on
SE/Security and found that the proposed inconsistencies have
not been covered before, while NFTGuard~\cite{nftdefects} and Pied-Piper~\cite{piedpiper} have similar detection patterns for contract semantics of UTS and UFF, respectively. We compare \textsc{HyperCode} with NFTGuard, as Piped-Piper is not fully open source (missing the fuzzing part). We randomly select 50 DApps from the large-scale dataset and collect their contract sourcecode (NFTGuard only supports sourcecode). The result shows that \textsc{Hyperion} finds 16 problematic \textit{UTS} contract behaviors with a precision of 75\%, while NFTGuard does not report any bugs. Details are provided in our repository.



\vspace{-0.2cm}
\section{Threats to Validity}
\vspace{-0.2cm}
\para{External Validity}
Our approach, employing the symbolic execution technique, faces challenges due to the increasing complexity of DApp contracts~\cite{nftdefects}. This complexity can lead to a path explosion issue, making it difficult to analyze deeper program points. Additionally, our dataflow analysis presents a limitation in accurately recovering dataflow through struct data types.
In description analysis, despite instructing-tuning the LLaMA2, we encounter instances of information misidentification. This issue partly stems from the inherent nature of LLMs, where subtle nuances in data or prompts can influence output accuracy. However, identifying specific error causes or model misinterpretations remains a challenge.
However, \textsc{Hyperion} can be improved by an instruction-tuning process with a larger labeled DApps dataset, and more rules to identify complicated data structures in the contract.

\para{Internal Validity}
Our manual labeling process for LLM construction and evaluation might introduce errors, particularly in differentiating between FNs and TPs. To mitigate this, we implement a double-check procedure and continuously update our dataset to ensure accuracy. All experimental results and evaluations are transparently shared in our repository.





\vspace{-0.2cm}
\section{Related Work}

\para{Inconsistency detection in DApps}
Several previous works focus on unexpected behaviors in DApps. For instance, DAppHunter~\cite{zhou2023dapphunter} assesses consistency among user intentions, blockchain wallet transactions, and contract behaviors, though it does not delve into contract bytecode or DApp descriptions analysis. VetSC~\cite{duan2022towards} detects discrepancies between contract bytecode and DApp category rules, deduced from the text around DApp buttons or widgets. 
Our approach, \textsc{Hyperion}, introduces a broader scope of inconsistency, bridging the gap between DApp descriptions and contract behaviors using natural language understanding and program analysis.

\para{Smart contract security issues detection}
Recently, many program analysis tools have been developed to focus on detecting security problems in smart contracts. 
Static analysis tools like~\cite{luu2016making, mythril, rao2012sailfish} are widely used for security issue detection. Dynamic testing tools~\cite{jiang2018contractfuzzer, nguyen2020sfuzz,piedpiper}, along with machine learning approaches~\cite{zhang2022reentrancy}, also contribute to this landscape.
However, to the best of our knowledge, \textsc{Hyperion} is the first tool incorporating both description analysis and contract semantics extraction to report newly found inconsistencies.

\para{Inconsistency detection of traditional applications}
There are some related works focusing on detecting trustworthiness and inconsistency issues in mobile~\cite{yu2018ppchecker,datausage,DiffDroid}
and web applications~\cite{ocariza2017detecting}.
However, the inexistence of a framework with rich semantics in the smart contract programming language poses unique challenges to contract bytecode analysis.

The immutable nature of blockchain transactions and financial attributes highlights the risks of DApp inconsistencies.
Our work specifically addresses this challenge and bridges the gap to enhance the DApp ecosystem's security, assisting developers, users, and marketplaces in identifying inconsistencies.

\vspace{-0.2cm}
\section{Conclusion}
In this paper, we define seven types of DApp inconsistencies from an empirical study and introduce \textsc{Hyperion}, a tool using LLM and dataflow-guided symbolic execution to identify inconsistencies between DApp descriptions and smart contract implementations automatically. \textsc{Hyperion} instruction-tunes LLaMA2 for DApp description analysis and utilizes dataflow-guided symbolic execution for contract bytecode analysis. The experimental results show our \textsc{Hyperion}'s effectiveness in unveiling DApp inconsistencies with an overall precision of 92.06\% and an overall recall of \OverallRecall.

\vspace{-0.1cm}
\section*{Acknowledgment}
This work is partially supported by fundings from the National Key R\&D Program of China (2022YFB2702203), the National Natural Science Foundation of China (62302534, 62332004).

\bibliographystyle{IEEEtran}
\bibliography{IEEEabrv,ref}

\end{document}